\documentclass[aps,prl,twocolumn,showpacs,amsmath,amssymb]{revtex4}
\usepackage{epsfig}
\usepackage{bm}
\usepackage{color}

\begin{document}

\title{
Noise and Full Counting Statistics of Incoherent Multiple Andreev Reflection  
}

\author{S. Pilgram and P. Samuelsson}

\affiliation{D\'epartement de Physique Th\'eorique, Universit\`e de Gen\'eve, CH-1211, Gen\`eve 4, Switzerland}

\date\today

\begin{abstract}
We present a general theory for the full counting statistics of
multiple Andreev reflections in incoherent
superconducting-normal-superconducting contacts. The theory, based on
a stochastic path integral approach, is applied to a
superconductor-double barrier system. It is found that all cumulants
of the current show a pronounced subharmonic gap structure at voltages
$V=2\Delta/en$. For low voltages $V\ll\Delta/e$, the counting
statistics results from diffusion of multiple charges in energy space,
giving the $p$th cumulant $\langle Q^p\rangle \propto V^{2-p}$,
diverging for $p\geq 3$. We show that this low-voltage result holds
for a large class of incoherent superconducting-normal-superconducting
contacts.

\end{abstract}

\maketitle

Electrical transport in superconducting contacts is a subject of
persistent interest. The basic mechanism of transport in voltage
biased contacts, Multiple Andreev Reflections (MAR), was described by
Klapwijk, Blonder and Tinkham\ \cite{KBT82} twenty years ago. Since
then, a large number of works have investigated the current-voltage
characteristics in superconducting contacts. As a prominent example,
experiments on superconducting atomic point contacts have been
explained remarkably well by coherent MAR-theories\ \cite{MARrew}.

Recently, interest has turned to the properties of current noise.
Shot noise was measured in several types of superconducting contacts\
\cite{Dieleman,Cron,LongDiff} and was theoretically studied in both the
coherent\ \cite{CohTheory1,CohTheory2} and incoherent limit\
\cite{LongTheoryBez1,LongTheoryNag,LongTheoryBez2}.  Very recently,
the full distribution of current fluctuations, the Full Counting
Statistics (FCS)\ \cite{Levitov93} of MAR in coherent superconducting
contacts was calculated\ \cite{Johansson03,Cuevas03}. The FCS revealed
explicitly that charge in coherent
superconducting-normal-superconducting (SNS) systems is transfered in
multiples of the elementary charge. In diffusive SNS-systems, this
leads to a characteristic power-law divergence of the shot noise at
low voltages, $S_I \sim V^{-1/2}$\ \cite{CohTheory2}. However, in many
experiments quantum coherence is suppressed\
\cite{LongDiff}. Interestingly, calculations for incoherent diffusive
structures\ \cite{LongTheoryBez1,LongTheoryNag,LongTheoryBez2} show
that the low-voltage shot noise is well-behaved and can be explained
in terms of an effective electron temperature of the order of the
superconducting gap $\Delta$. This naturally raises the question about
the role of multiple charge transfer in incoherent SNS-systems.

We address this question by developing a theory for the FCS of
incoherent MAR in semiclassical SNS-systems, based on a stochastic
path integral\ \cite{Pilgram03,Jordan04} approach. As an illustrative
example, we investigate the superconducting double-barrier model of
Octavio et al. (OTBK)\ \cite{OBTK83}, focusing on the voltage
dependence of the FCS. For a generic incoherent SNS-system, we show
that for low voltage, $eV\ll \Delta$, the generating function is
\begin{equation}
S(\chi)=\frac{\tau eV^2G}{2\Delta}\mbox{asinh}^2\sqrt{\mbox{exp}(\chi
2\Delta/eV)-1},
\label{Low eV action 3}
\end{equation}
with $\tau$ the measurement time, $\chi$ the counting field and $G$
the linear, low-voltage conductance. This result, an effect of
diffusive motion of the quasiparticles in energy space, is just the
FCS for a metallic diffusive wire\ \cite{Jordan04,Lee95} with an
effective voltage dependent charge $e\rightarrow
e(2\Delta/eV)$. Unlike coherent diffusive systems\ \cite{CohTheory2},
the cumulants $\langle Q^p \rangle =\partial^p S/\partial
\chi^p|_{\chi=0}\sim V^{2-p}$ diverge only for $p \geq 3$. The
successful measurement of the third cumulant $\langle Q^3 \rangle$ in
normal conducting tunnel junctions\ \cite{Reulet03} shows that an
experimental test of our prediction is feasible.

\begin{figure}[b]
\epsfxsize8cm \centerline{\epsffile{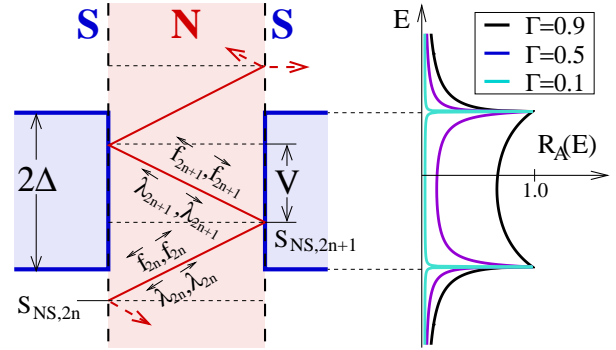}}
\vspace{5mm}
\caption{ Left panel: A schematic of the geometry and the MAR-ladder in energy
space (see text for details). Right panel: Energy dependent probability
$R_A(E)$ of a single Andreev reflection for different interface transparencies
$\Gamma$.}
\label{Geometry}
\end{figure}
\noindent
{\it The model}\ --\ The studied geometry is shown in Fig.\
\ref{Geometry}. Two superconducting electrodes (S) with gaps $\Delta$
are connected via a ballistic normal conducting region (N) with $M$
transport modes.  The NS-interfaces, with equal mode-independent
normal state transparencies $\Gamma$, are characterized by energy
dependent probabilities\ \cite{BTK83} for normal/Andreev reflection
and transmission $R_{N/A}(E),T_{N/A}(E)$. A voltage $V$ is applied
between the two superconductors. For $eV<2\Delta$, transport of
quasiparticles across the gap takes place via MAR along the energy
ladder shown in Fig.\ \ref{Geometry}.

In contrast to previous works on FCS of MAR
\cite{Cuevas03,Johansson03}, we will consider the incoherent limit,
i.e. with suppressed proximity and dc/ac-Josephson effect. This
experimentally interesting limit\ \cite{LongDiff} is relevant for
strong dephasing in the normal region (for incoherent transport in
diffusive NS-systems, see Refs.\ \cite{Belzig04}).  We assume low
temperature, $kT\ll\Delta$, and negligible relaxation due to
electron-electron or electron-phonon scattering. Furthermore, we focus on
semiclassical systems with $M\gg 1$.

In this situation, following OTBK\ \cite{OBTK83}, the state of the
normal region is described by energy, space and direction dependent
occupation functions $\overleftarrow{f_n},\overrightarrow{f_n}$ for
left- and right-going particles ($n$ even) and holes ($n$ odd) [see
Fig.\ \ref{Geometry}]. The average current through the junction is
obtained from a set of coupled linear equations, the OBTK-equations,
for the occupation functions on different rungs of the energy
ladder. In the present work, we show how to extend the OBTK-equations
to determine the current noise and the FCS in the system. We however
emphasize that incoherent diffusive\
\cite{LongTheoryBez1,LongTheoryNag,LongTheoryBez2} or chaotic
contacts\ \cite{Samuelsson02} can be treated in a very similar
fashion. Below we put $e=\hbar=1$.

Introduced by Levitov and Lesovik\ \cite{Levitov93}, the FCS
\begin{equation}
P(Q) = \frac{1}{2\pi} \int d\chi e^{-i\chi Q}e^{S(i\chi)},\quad Q =
\int_0^\tau dt I(t),
\label{Full Statistics Definition}
\end{equation}
is the distribution giving the probability that a certain charge $Q$
flows through an electrical conductor during the time interval
$[0,\tau]$. It describes entirely the low-frequency current
fluctuations, if $\tau$ is chosen longer than any intrinsic time
scale. The Fourier transform of the probability distribution $P(Q)$
yields the generating function $S(\chi) = \sum \chi^p \langle Q^p
\rangle/p!$ of irreducible cumulants and gives the mean
current $ I = \langle Q \rangle / \tau$, the current noise $S_I =
\langle Q^2 \rangle / \tau$, the third cumulant $C_3=\langle Q^3
\rangle/\tau$ describing the asymmetry of the distribution, etc.

{\it Formal Solution}\ --\ The FCS of a single NS-interface is a
multinomial process and can be fully described\ \cite{Muzykantskii94}
by the reflection and transmission probabilities
$R_{N/A},T_{N/A}$. Inside the gap, $|E| < \Delta$, the generating
function is
$$
S_{\text{NS}} = \frac{M\tau}{2\pi} 
\ln \Bigl\{
f_p f_h + f_p(1-f_h) (R_A e^{2\chi} + R_N)
$$
\vspace{-5mm}
\begin{equation}
 + f_h(1-f_p) (R_A e^{-2\chi} + R_N)
+ (1-f_p)(1-f_h)
\Bigr\} 
\label{NS-Interface}
\end{equation}
where $f_p,f_h$ are the particle and hole occupation functions in the
normal region. A similar expression accounts for reflection and
transmission outside the gap, $|E| > \Delta$. The SNS-junction can be
considered as a series of two NS-interfaces sharing the same
occupation functions. In the incoherent semiclassical limit, the FCS
of such a compound system can be calculated with the stochastic path
integral formalism\ \cite{Pilgram03,Jordan04}. The key ingredient in
this formalism is a separation of time scales for fast collision
events (the Andreev reflections) and for the slow evolution of the
occupation functions. This separation allows one to integrate out the
fast fluctuations and to express the generating function $S(\chi)$ by
an action in terms of slow collective variables as,
\begin{equation}
S[\chi,f,\lambda] = \int_0^{2V} dE \sum_{n=-\infty}^{\infty}
S_{\text{NS},n} (E_n),
\label{Full Action}
\end{equation}
where the integral is over the different ladders, contributing
independently to the action. The actions $S_{\text{NS},n}$ evaluated
at $E_n =E +nV$ (see Fig.\ \ref{Geometry}) are generalizations of Eq.\
(\ref{NS-Interface}) which distinguish left- and right-going
currents. E.g. for the left interface ($n$ even), for
$|E_n|<\Delta$ we find
$$
S_{\text{NS},n}(E_n)
= \frac{M\tau}{2\pi}  \ln
\Bigl\{
\overleftarrow{f_{n}}_{-1}
\overleftarrow{f_n}
e^{\overrightarrow{\lambda_{n}}_{-1}+\overrightarrow{\lambda_n}
-\overleftarrow{\lambda_{n}}_{-1}-\overleftarrow{\lambda_n}}
+
$$
\vspace{-5mm}
$$
\overleftarrow{f_{n}}_{-1}
(1-\overleftarrow{f_n})
\left(
R_{A} e^{2\chi+\overrightarrow{\lambda_n}
-\overleftarrow{\lambda_{n}}_{-1}}
+
R_{N} e^{\overrightarrow{\lambda_{n}}_{-1}
-\overleftarrow{\lambda_n}_{-1}}
\right)
+
$$
\vspace{-5mm}
$$
\overleftarrow{f_n}
(1-\overleftarrow{f_n}_{-1})
\left(
R_{A} e^{-2\chi+\overrightarrow{\lambda_n}_{-1}
-\overleftarrow{\lambda_n}}
+
R_{N} e^{\overrightarrow{\lambda_n}
-\overleftarrow{\lambda_n}}
\right)
$$
\vspace{-5mm}
\begin{equation}
+
(1-\overleftarrow{f_n}_{-1})
(1-\overleftarrow{f_n})
\Bigr\}.
\label{Generator Example}
\end{equation}
In addition to the occupation functions
$\overleftarrow{f_n},\overrightarrow{f_n}$ we introduce internal
counting fields $\overleftarrow{\lambda_n},\overrightarrow{\lambda_n}$
which are Lagrange multipliers preventing charge accumulation inside
the normal region. The action\ (\ref{Full Action}) has to be varied
over all possible configurations of $f_n$'s and $\lambda_n$'s\
\cite{Pilgram03}. In the semiclassical regime, we may calculate the
action in the saddle point approximation using the equations of
motion\ \cite{Pilgram03},
\begin{equation}
\frac{\partial S}{\partial \overleftarrow{\lambda}_n} 
=
\frac{\partial S}{\partial \overrightarrow{\lambda}_n} 
=
\frac{\partial S}{\partial \overleftarrow{f}_n} 
=
\frac{\partial S}{\partial \overrightarrow{f}_n} = 0.
\label{Saddle Point Equations}
\end{equation}
These equations form an infinite system of coupled nonlinear equations
which have to be solved for arbitrary $\chi$ [the occupation of
incoming quasiparticles is unity (zero) at $E<-\Delta$
$(E>\Delta)$]. The solutions are then substituted back into Eq.\
(\ref{Full Action}) to obtain $S(\chi)$.  The Fourier transform\
(\ref{Full Statistics Definition}) yielding the probability
distribution $P(Q)$ can be carried out in the stationary phase
approximation. The relation to the OBTK equations becomes clear if we
calculate the average current $I= dS(\chi)/d\chi|_{\chi=0}/\tau$. For
this purpose, it is sufficient to solve the saddle point equations for
$\chi=0$, giving
$\overleftarrow{\lambda}_{n},\overrightarrow{\lambda}_{n}=0$. The
first two derivatives in Eq.\ (\ref{Saddle Point Equations}) then
become exactly the OBTK equations for the occupation functions. For
the noise $S_I=d^2S(\chi)/d\chi^2|_{\chi=0}/\tau$, this procedure is
no longer adequate and the saddle point equations then have to be
solved to first order in $\chi$.

\begin{figure}[t]
\epsfxsize8cm \centerline{\epsffile{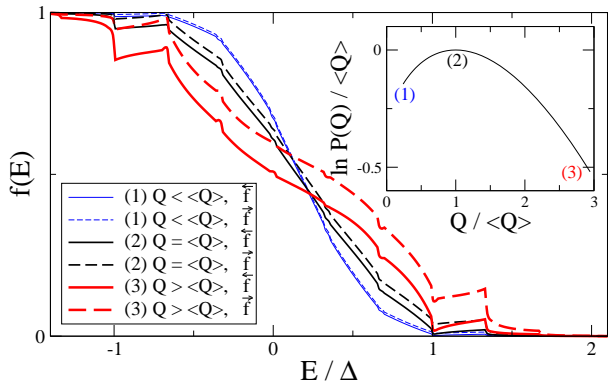}}
\vspace{5mm}
\caption{The FCS for $V=0.34\Delta$ and $\Gamma=0.5$. Main panel: The
  left- and rightgoing distribution functions at the right interface
  for a transmitted charge $Q<\langle Q \rangle$, $Q=\langle Q
  \rangle$ and $Q>\langle Q \rangle$. Upper inset: The probability
  distribution $P(Q)$ of transmitted charge.}
\label{Saddle Point Solutions}
\end{figure}
\noindent

{\it Numerical Results}\ --\ In general, the FCS has to be calculated
numerically. A typical distribution function $P(Q)$ is shown in Fig.\
\ref{Saddle Point Solutions}: the FCS is asymmetric, bounded from the left
side, $Q>0$ (since charge may not flow against the bias) and exhibits a long
tail on the right side (since the overall probability to undergo MAR and to
climb up the energy ladder is small). Saddle point solutions for the $f$'s
belonging to different points of the probability distribution show
characteristic structures at $E=\pm\Delta + nV$ due to the cusps in the
Andreev reflection probability (see Fig.\ \ref{Geometry}). The occupation
functions $\overrightarrow{f},\overleftarrow{f}$ can be interpreted as the
most probable configurations under the condition that a certain charge
$Q\propto\int dE (\overrightarrow{f}-\overleftarrow{f})$ is transmitted. If
the output is the mean charge $Q=\langle Q \rangle$, the solution is exactly
the one found by the OBTK equations. For a small amount of transmitted charge
$Q<\langle Q \rangle$, the difference $\overrightarrow{f}-\overleftarrow{f}$
becomes tiny (no net current).  In the opposite limit $Q$ $>$ $\langle
Q\rangle$, this difference is large.

\begin{figure}[t]
\epsfxsize8cm
\centerline{\epsffile{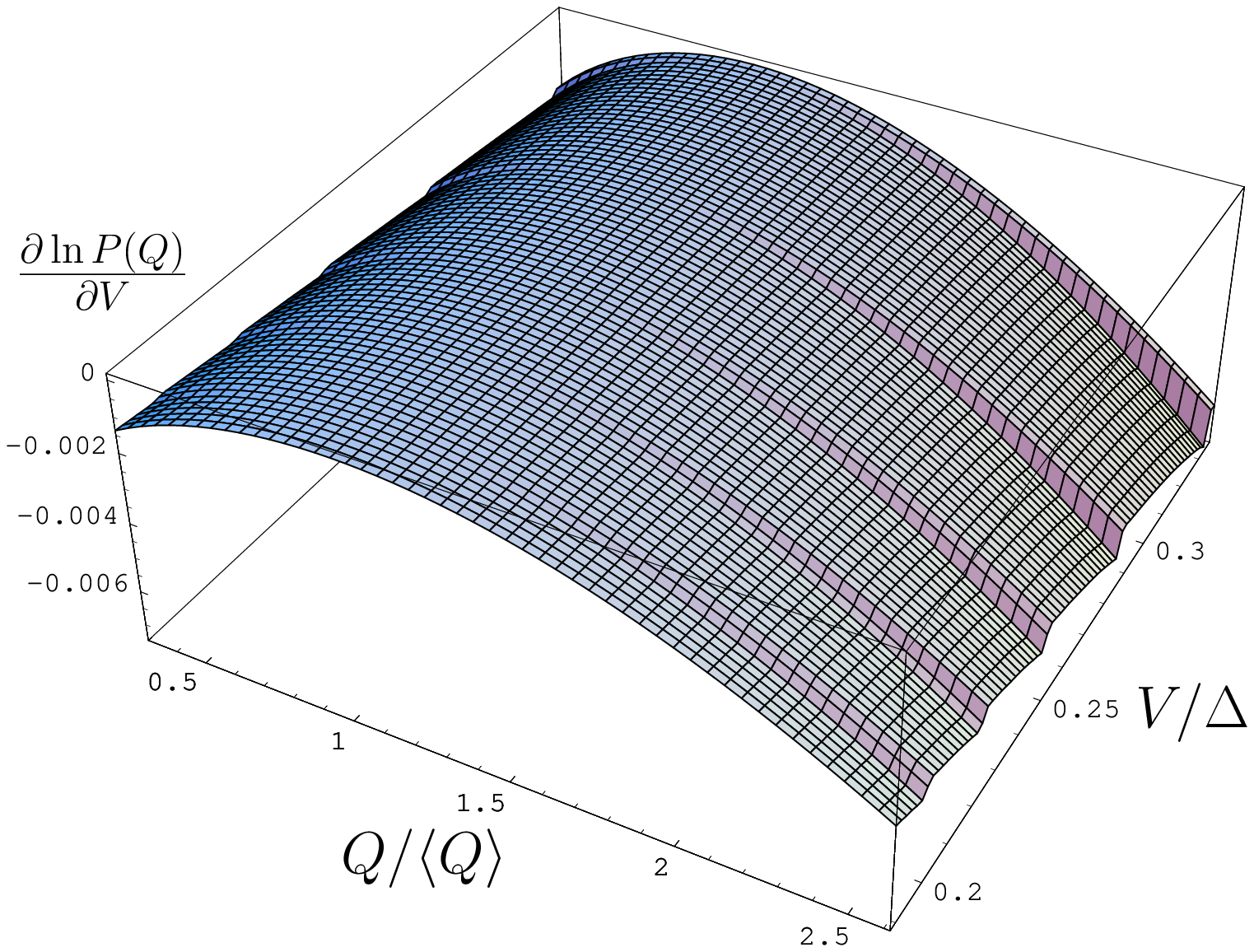}}
\vspace{5mm}
\caption{Differential full counting statistics as function of the
  voltage for the MAR-system in the tunneling limit, $\Gamma=0.1$.}
\label{Full Statistics}
\end{figure}

\begin{figure}[t]
\epsfxsize8cm \centerline{\epsffile{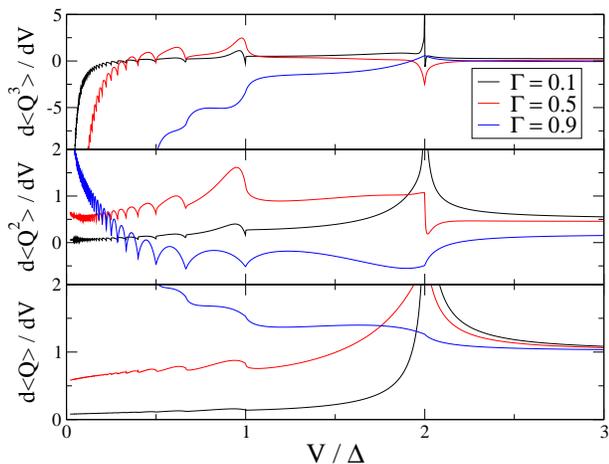}}
\vspace{5mm}
\caption{The first three differential cumulants of the FCS, normalized
  with the normal state conductance, as a function of voltage for
  three different transparencies $\Gamma$ of the NS-interfaces.}
\label{Cumulants}
\end{figure}

The differential FCS, shown in Fig.\ \ref{Full Statistics}, show a
pronounced subharmonic gap structure, with cusps at voltages
$V=2\Delta/n$. Note that the derivative of the probability
distribution with respect to voltage is always negative, i.e. the
distribution gets broader towards the low voltage limit and acquires a
strong tail for $Q>\langle Q\rangle$. The first three differential
cumulants $d\langle Q^n \rangle / dV$ are plotted in
Fig.\ \ref{Cumulants} for transparencies $\Gamma=0.1, 0.5$ and
$0.9$. All cumulants show a subharmonic gap structure, more pronounced
for the noise and third cumulant than for the current \cite{Klapwijk}.
\noindent

{\it Voltage limits}\ --\ For high voltages $V \gg \Delta$, only
quasiparticles traversing the junction once contribute to the
differential transport statistics. The saddle point equations in\
(\ref{Saddle Point Equations}) can then be solved analytically, giving
the FCS [$q=Q/(MV\tau/(2\pi))$]
\begin{equation}
\frac{\partial \ln P(Q)}{\partial V} = \frac{M\tau}{2\pi} 
\left\{\ln\frac{(1-\Gamma)(1+q)}{1-q} -
q\ln\frac{4(1-\Gamma)q^2}{1-q^2}\right\},
\label{nbarr}
\end{equation}
the result for the double barrier system in Fig.\ \ref{Geometry} in
the normal state \cite{deJong96}. Unlike the numerical result for
intermediate voltages in Fig.\ \ref{Full Statistics}, this high
voltage result scales linearly with $V$ and is bounded from above
$(0<q<1)$.

For low voltage $V \ll \Delta$, the quasiparticles injected below the gap
traverse the junction a large number of times $N=2\Delta/V\gg1$ before being
emitted above the gap. For sufficiently strong normal backscattering at the
NS-interfaces, $\Gamma^N\ll \Gamma/N$, the motion in energy space, along the
ladder in Fig.\ \ref{Geometry}, becomes diffusive\ \cite{LongTheoryNag}.  We
may introduce a scale $m$ with $1\ll m \ll N$ below which the motion is
ballistic and described by an energy-dependent conductance per rung,
$\tilde{G}=MR_A(E)/(2\pi[1-R_A(E)])$. On larger scales, one can apply a
diffusion approximation $\overleftarrow{f_n} \simeq \overrightarrow{f_n}
\equiv f_n$ and $f_{n+m} \simeq f_n + m(\partial f_n / \partial n)$ (and
similarly for the $\lambda$'s) to Eq.\ (\ref{Generator Example}).  Replacing
the sum in Eq.\ (\ref{Full Action}) by an integral (putting $dn= dE/V$), we
obtain in analogy to the procedure for a normal diffusive wire\
\cite{Jordan04} the action
\begin{equation}
\begin{array}{c}
\frac{S}{2\tau V^2}=
\int\limits_{-\Delta}^{\Delta}dE~\tilde{G}\left[f(1-f)\left(\frac{\partial \lambda
}{\partial E}+\frac{\chi}{V}\right)^2-\frac{\partial f}{\partial E} \left(\frac{\partial \lambda}{\partial
E}+\frac{\chi}{V}\right)\right],
\end{array}
\label{Low eV action 2}
\end{equation}
Changing variables $\lambda \mapsto \lambda + \chi E/V$ and $dE\sim
\tilde{G} dy$ maps Eq.\ (\ref{Low eV action 2}) onto the action of a
normal diffusive wire\ \cite{Jordan04} with renormalized, voltage
dependent charge $e \rightarrow e(2\Delta/V)$.  This gives directly
the generating function in Eq.\ (\ref{Low eV action 3}) with
$G=\Delta[\int_{-\Delta}^{\Delta} dE~\tilde{G}^{-1}]^{-1}=
3M\Gamma^2/(2\pi[16(1-\Gamma)])$.
\noindent
Consequently, for low voltages, the current $I =VG$ is proportional
to voltage, while the noise $S_I=4\Delta G/3$ saturates a a constant
value. All higher cumulants $p\geq 3$ diverge as $V^{2-p}$.

{\it Generalization}\ --\ Importantly, the stochastic path integral
method presented here can be employed to any semiclassical, incoherent
mesoscopic SNS-junction. In particular, Eq.\ (\ref{Low eV action 3})
gives the low voltage FCS of any SNS-junction with sufficient normal
back-scattering. We emphasize that the FCS is characterized by a
single parameter, the low-voltage conductance $G$.  For a diffusive
normal region with conductance $G_W$ we find for instance $G =
[4\pi(8(1-\Gamma) + 3\Gamma^2)/(3M\Gamma^2) + 2/G_W]^{-1}$. For
negligible interface resistance, $M/(2\pi) \gg G_W$, Eq.\ (\ref{Low eV
action 3}) then extends the shot-noise result of Refs.\
\cite{LongTheoryNag,LongTheoryBez2} to the FCS. The opposite limit
$M/(2\pi) \ll G_W$ corresponds to an incoherent chaotic cavity. One
should however keep in mind that eventually, for sufficiently low
voltage, inelastic electron-electron and electron-phonon scattering
will dominate the transport and the low-voltage result in Eq.\
(\ref{Low eV action 3}) will not apply.

To access the FCS for the full voltage range, the numerical scheme
presented for the OTBK-system has to be appropriately modified. E.g,
for the normal region being a chaotic cavity, the only modification is
to consider an isotropic occupation function
$\overleftarrow{f_n}=\overrightarrow{f_n} = f_n$ (and
$\overleftarrow{\lambda_n}=\overrightarrow{\lambda_n} = \lambda_n$) in
Eq.\ (\ref{Generator Example}). For a diffusive SNS-junction, one can
directly extend the semiclassical approach for current and noise in
Refs.\ \cite{LongTheoryBez1,LongTheoryNag,LongTheoryBez2} to the
FCS. At a given voltage $2\Delta/(n+1)<V<2\Delta/n$, quasiparticles
are transfered through the gap via $n+1$ or $n+2$ traversals across
the junction, transporting $n+1$ or $n+2$ charges respectively. The
transport is effectively through a diffusive wire of conductance
$G_W/(n+1)$ or $G_W/(n+2)$. The action is then just the sum of the
actions of these two diffusive transport processes, with appropriate
weights,
\begin{eqnarray}
&&S/(G_W\tau)=\frac{2\Delta-nV}{n+2}\mbox{asinh}^2\sqrt{\mbox{exp}[\chi (n+2)]-1} \nonumber \\
&+&\frac{(n+1)V-2\Delta}{n+1}\mbox{asinh}^2\sqrt{\mbox{exp}[\chi
(n+1)]-1}.
\label{diffusive act}
\end{eqnarray}
The two first cumulants\ \cite{LongTheoryNag,LongTheoryBez2}
$I=G_WV$ and $S_I=2G_W(2\Delta+V)/3$ are smooth functions of
voltage. However, higher cumulants exhibit kinks at $V=2\Delta/n$.

In conclusion, we have presented a theory for the full counting
statistics of incoherent multiple Andreev reflection, based on the
stochastic path integral approach. The charge is transfered in quantas
of multiple electron charge, giving rise to a low-voltage divergence
for cumulants of order three and higher.

This work was supported by the Swiss NSF and the program MaNEP.


\begin{thebibliography}{99}
\bibitem{KBT82}
T.~M.~Klapwijk, G.~E.~Blonder, and M.~Tinkham, Physica B+C {\bf
  109-110}, 1657 (1982).
\bibitem{MARrew}
 N. Agrait, A. Levy Yeyati, J. M. van Ruitenbeek, Phys. Rep. {\bf
 377}, 81 (2003).
\bibitem{Dieleman}
P. Dieleman {\it et al.}, Phys. Rev. Lett. {\bf 79}, 3486 (1997).
\bibitem{Cron} 
R. Cron {\it et al.}, Phys. Rev. Lett.  {\bf 86}, 4104 (2001);
F.E. Camino, cond-mat/0406650.
\bibitem{LongDiff} 
X. Jehl {\it et al.}, Phys. Rev. Lett. {\bf 83},
 1660 (1999); T. Hoss {\it et al.}, Phys. Rev. B {\bf 62}, 4079
  (2000); P. Roche {\it et al.}, Physica C {\bf 352}, 73 (2001); C. Hoffmann, F. Lefloch, and M. Sanquer, Euro. Phys. J B
  {\bf 29}, 629 (2002).
\bibitem{CohTheory1} 
J.P. Hessling {\it et al}, Europhys. Lett. {\bf
34}, 49 (1996); D. Averin and H.T. Imam, Phys. Rev. Lett. {\bf 76},
3814 (1996); J.C. Cuevas, A. Martin-Rodero, and A. Levy-Yeyati, {\it
  ibid} {\bf 82}, 4086 (1999).
\bibitem{CohTheory2}
Y. Naveh and D.V. Averin, Phys. Rev. Lett. {\bf 82}, 4090 (1999).
\bibitem{LongTheoryBez1}
E. V. Bezuglyi {\it et al.}, Phys. Rev. Lett. {\bf 83}, 2050 (1999).
\bibitem{LongTheoryNag} 
K. E. Nagaev, Phys. Rev. Lett. {\bf 86}, 3112 (2001). 
\bibitem{LongTheoryBez2}
E. V. Bezuglyi {\it et al.}  Phys. Rev. B {\bf 63} 100501
(2001).
\bibitem{Levitov93} L. S. Levitov and G. B. Lesovik,
JETP Lett.\ {\bf  58}, 230
(1993); cond-mat/9401004; L. S. Levitov, H. Lee,
and G. B. Lesovik, J. Math.\
Phys.\ {\bf  37}, 4845 (1996).
\bibitem{Johansson03}
G.~Johansson, P.~Samuelsson, \AA.~Ingerman, Phys. Rev. Lett. {\bf 91}, 187002 (2003).
\bibitem{Cuevas03}
J.~C.~Cuevas and W.~Belzig, Phys. Rev. Lett. {\bf 91}, 187001 (2003); condmat/0406508.
 \bibitem{Pilgram03}
S.~Pilgram, A.~N.~Jordan, E.~V.~Sukhorukov, and M.~B\"uttiker,
Phys. Rev. Lett. {\bf 90}, 206801 (2003).
%S.~Pilgram {\it et al.}, Phys. Rev. Lett. {\bf 90}, 206801 (2003).
\bibitem{Jordan04}
A.~N.~Jordan, E.~V.~Sukhorukov, and S.~Pilgram, 
 cond-mat/0401650.
\bibitem{OBTK83}
M.~Octavio {\it et al.}, Phys. Rev. B {\bf 27}, 6739 (1983).
\bibitem{Lee95}
H.~Lee, L.~S.~Levitov, A.~Yu.~Yakovets,
Phys. Rev. B {\bf 51}, 4079 (1995); Yu.~V.~Nazarov,
Ann. Phys. (Leipzig) {\bf 8} Spec. Issue, S1-193 (1999); T.~Bodineau and B.~Derrida, 
Phys. Rev. Lett. {\bf 92}, 180601 (2004); D.~B.~Gutman, A.~D.~Mirlin,
and Y.~Gefen, cond-mat/0403436 (2004).
\bibitem{Reulet03}
B.~Reulet, J.~Senzier, and D.~E.~Prober, 
Phys. Rev. Lett. {\bf 91}, 196601 (2003).
\bibitem{BTK83} 
G.~E.~Blonder, M.~Tinkham, and T.~M.~Klapwijk, Phys. Rev. B {\bf 27},
112 (1983). 
\bibitem{Belzig04} 
E. V. Bezuglyi {\it et al.}, Phys. Rev. B {\bf 62}, 14439 (2000);
K. Nagaev and M. B\"uttiker, {\it ibid} {\bf 63}, 081301 (2001);
W. Belzig and P. Samuelsson, Europhys. Lett. {\bf 64}, 253 (2003).
\bibitem{Samuelsson02}
P.~Samuelsson {\it et al}, Phys. Rev. B  {\bf 65}, 180514 (2002).
\bibitem{Muzykantskii94}
B. A. Muzykantskii and D. E. Khmelnitskii,
Phys. Rev. B {\bf 50}, 3982 (1994); W.~Belzig and Yu.~V.~Nazarov, Phys. Rev. Lett. {\bf 87}, 197006 (2001). 
\bibitem{Klapwijk}
The semiclassical noise-theory of Ref.\ \cite{Dieleman} gives a 
qualitatively similar but quantitatively different result.
\bibitem{deJong96}
M.~J.~M.~de Jong, Phys. Rev. B {\bf 54}, 8144 (1996); P.-E.~Roche, B.~Der\-ri\-da, and B.~Dou\-\c{c}ot,
cond-mat/0312659.

\end{thebibliography}
\end{document}